# Cascaded Generation of a Sub-10-Attosecond Half-Cycle Pulse

Yinren Shou[1], Ronghao Hu[2], Zheng Gong[1], Jinqing Yu[3], Jia erh Chen[1], Gerard Mourou[4], Xueqing Yan[1,5,6,7], Wenjun Ma[1,6,7,*]

[1]State Key Laboratory of Nuclear Physics and Technology, and Key Laboratory of HEDP of the Ministry of Education, CAPT, School of Physics, Peking University, 100871 Beijing, China

[2]College of Physics, Sichuan University, 610065 Sichuan, China

[3]School of Physics and Electronics, Hunan University, 410082 Hunan, China

[4]DER-IZEST, Ecole Polytechnique, 91128 Palaiseau Cedex, France

[5]Collaborative Innovation Center of Extreme Optics, Shanxi University, 030006 Shanxi, China

[6]Beijing Laser Acceleration Innovation Center, Huairou, 101400 Beijing, China

[7]Institute of Guangdong Laser Plasma Technology, Baiyun, 510540 Guangzhou, China

[*]Corresponding authors. Email and telephone number: wenjun.ma@pku.edu.cn

## Abstract

Sub-10-attosecond pulses with half-cycle electric fields provide exceptional options to detect and manipulate electrons in the atomic timescale. However, the availability of such pulses is still challenging. Here, we propose a method to generate isolated sub-10-attosecond half-cycle pulses based on a cascade process naturally happening in plasma. A 100s-attosecond pulse is first generated by shooting a moderate overdense plasma with a one-cycle femtosecond pulse. After that, the generated attosecond pulse cascadedly produce a sub-10-attosecond half-cycle pulse in the transmission direction by unipolarly perturbing a nanometer-thin relativistic electron sheet naturally form in the plasma. Two-dimensional particle-in-cell simulations indicate that an isolated half-cycle pulse with the duration of 8.3 attoseconds can be produced. Apart from one-cycle driving pulse, such a scheme also can be realized with a commercial 100-TW 25-fs driving laser by shaping the pulse with a relativistic plasma lens in advance.

## Introduction

Half-cycle pulses, also called unipolar pulses, refer temporally asymmetrical near-single-cycle pulses whose fields in one polarity are predominantly stronger than that in the opposite polarity[1]. Such pulses normally generate from collective unidirectional motion of electrons. In the near-field, half-cycle pulses can offer exceptional possibilities to control and probe electron dynamics by asymmetrically manipulating electron wavepackets in atoms or in solid[2,3]. In the far-field, the asymmetrical feather can be maintained with a long but much weaker tail ensuring the integral of the field is zero, which is still very valuable for many applications not requiring the strict unipolarity[4]. In experiments, half-cycle pulses have been obtained in the terahertz[5], far-infrared[6], mid-infrared[7], and visible ranges[8]. Recently, the generation of attosecond half-cycle pulses in extreme-ultra-violet (XUV) range from relativistic laser-plasma interaction[9-11] is attracting much attention. Such attosecond XUV pulses would bring new opportunities to several applications such as nonlinear optics in the XUV region[12], XUV pump-probe spectroscopy of ultrafast dynamics[13], and single-shot diffractive imaging of a single biomolecule[14].

Several schemes for the production of isolated half-cycle attosecond pulses from solid surfaces or underdense plasma have been studied. Naumova et al. proposed that a single 200-attosecond pulse could be produced when a tightly focused laser pulse is reflected from a relativistic plasma mirror created by the pulse itself[15]. Wu et al. numerically demonstrated that an intense half-cycle pulse with a duration of tens of attoseconds could be produced by irradiating a double-foil target with a few-cycle laser pulse[16]. Continuum XUV spectra that support a 600-attosecond pulse have been experimentally observed in this scheme, but the temporal profile of the pulse still needs to be measured[17]. Attosecond half-cycle pulses generated from underdense plasma also have been studied by Li et al[18]. They proposed that an intense, radially polarized, 100-attosecond half-cycle pulse can be generated from ultrathin relativistic electron disks in a quasi-one-dimensional wakefield acceleration regime[18]. Further reduction in durations of the attosecond pulses can improve the ultimate temporal resolutions in experiments. However, the direct generation of a brilliant half-cycle pulse with sub-10-attosecond duration is still a challenge.

In the present work, we propose a theoretical scheme of generating an isolated sub-10-attosecond half-cycle pulse from the interaction of a relativistic laser pulse with a moderate overdense plasma. We first illustrate the emission mechanism and electron dynamics through one-dimensional (1D) particle-in-cell (PIC) simulations. An isolated, half-cycle, 3-attosecond (defined as the full width at half maximum of the intensity) pulse with a peak electric field of $10^{13}$ V/m is generated by utilizing a one-cycle driving pulse. Second, the robustness of our scheme is testified by varying the densities of the plasmas and the carrier-envelope phases (CEPs) of the driving pulses. In 2D simulations, the half-cycle feature maintains, and the pulse duration increases to 8.3 attoseconds with the similar parameters. Finally, we demonstrate that our scheme can be realized presently with a 100-TW 25-fs driving pulse by placing a piece of near-critical-density (NCD) plasma slab in front of the overdense plasma to shape the pulse. After the shaping process, the multi-cycle pulse transforms into a pulse with a sharp rising edge, which leads to the generation of a 40-attosecond half-cycle pulse in the same scheme as the one-cycle driving laser.

## Results

**Emission mechanism of the half-cycle attosecond pulse**

The relativistic laser-plasma interactions are extremely nonlinear and complicated due to the strong coupling between the laser field and the self-induced field in palsma[19]. We first carried out 1D PIC simulations to illustrate the emission mechanism of the half-cycle attosecond pulse in our scheme. For the simplicity of the physics, a one-cycle laser pulse (see profile in Methods and Fig. s1 of Supplementary Information) is chosen as the driving pulse irradiated on a plasma with a moderate overdense electron density of 40 $n_c$. Here $n_c = m_e \epsilon_0 \omega_0^2 / e^2$ is the critical density of plasma, $m_e$ and $e$ is the rest mass and charge of an electron, $\omega_0$ is the laser angular frequency, and $\epsilon_0$ is the dielectric constant. The normalized amplitude of the driving pulse is a = 30, where a = $eE/m_e c \omega_0$, E is the peak electric field of the pulse and $c$ is the light speed in vacuum. With the developments of laser technologies, similar near-single-cycle laser pulses have been experimentally realized at non-relativistic[20-22] and moderately relativistic intensities[23,24]. Relativistic one-cycle pulses are also promising to be obtained from plasma optic devices[25,26]. It should be noted that a multi-cycle pulse with a sharp rising edge is also applicable for driving the cascade mechanism.

The generation process of the half-cycle attosecond pulse is depicted in Fig. 1 based on the simulative spatial-temporal evolution of the electron density. When the driving pulse irradiates the plasma, electrons are piled up by the ponderomotive force in the first half cycle of the laser[27], as shown in Fig. 1a). Consequently, an electrostatic field as strong as $10^{14}$ V/m is created due to the charge separation. When the laser field starts to decline after the peak of the first half cycle, the electrons are pulled back by the electrostatic restoring force. Then an electron sheet (ES1) is formed and backwardly accelerated to the relativistic speed inside the plasma, which simultaneously leads to the generation of a backward attosecond pulse (AP1) as depicted in Fig. 1b) at the simulation time of 6.8 fs. After ES1 reaches the left boundary of the plasma at the simulation time of 7.2 fs, the charge separation field reverses its direction. Driven by this field, part of the electrons in ES1 forms a secondary nanometer-thin electron sheet (ES2) moving in the forward (transmission) direction as displayed in Fig. 1c). In the strong charge separation field, ES2 is immediately accelerated to the speed of light, and meanwhile, transversely perturbed by AP1 in a constant direction. As a result of this transient perturbation, an isolated half-cycle pulse (AP2) propagating in the forward direction is produced as depicted in Fig. 1d). After that, ES2 is quickly dispersed by the deceleration field at $x > 1.2 \lambda_0$, whereas AP2 propagates out of the plasma maintaining the duration around 3 attoseconds as shown in Fig. 3a).

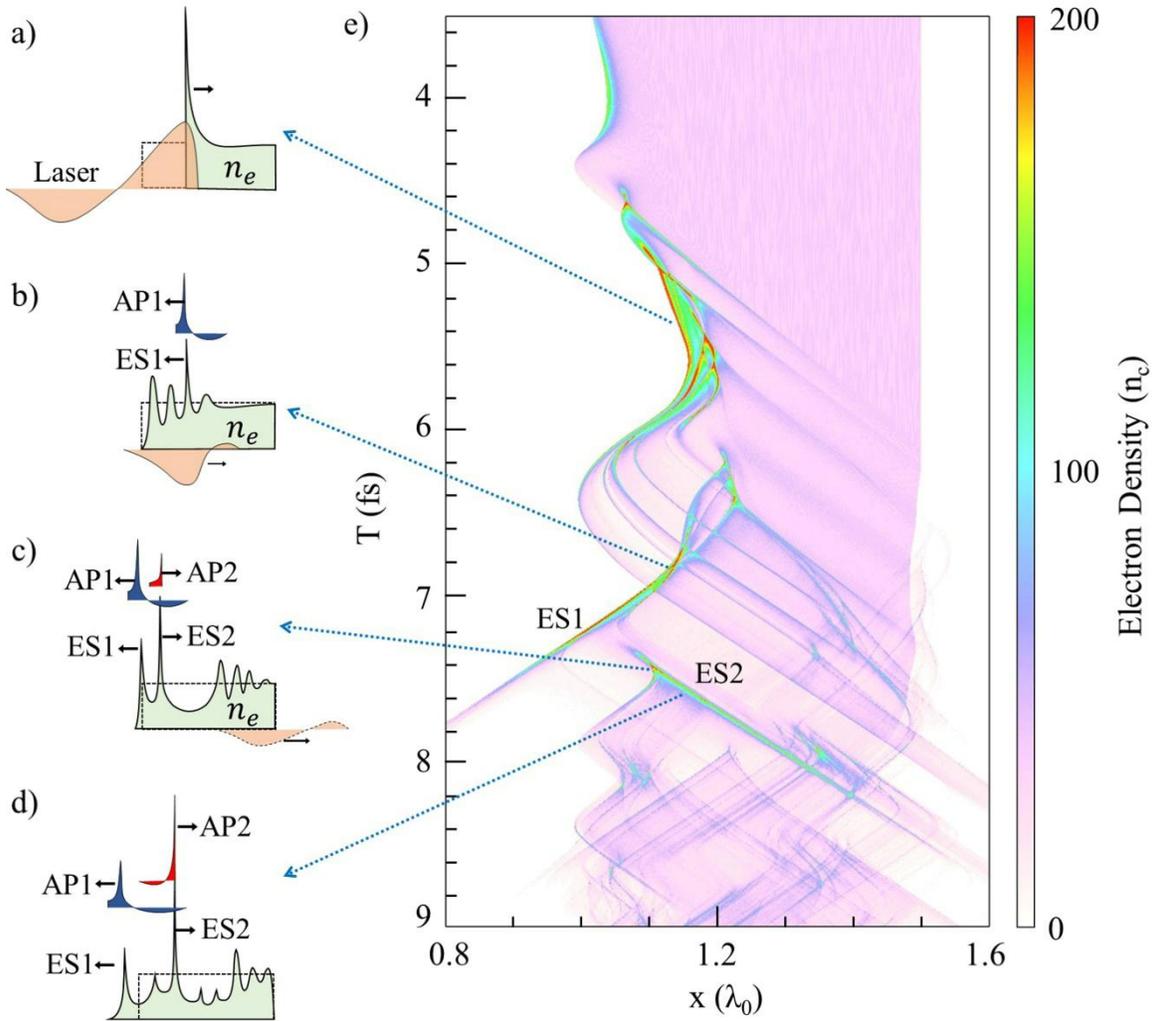

**Figure 1: The generation mechanism of a sub-10-attosecond half-cycle pulse.** a) - d) conceptual graphs derived from the simulations displaying the formations and interactions of the electron sheets (ES) and the attosecond pulses (AP) step by step as illustrated in the main text. e) The spatial-temporal evolution of the electron density n(x, t) in the unit of $n_c$.

The emission process of the half-cycle pulse is determined by the electron dynamics in ES2. Figure 2a) and 2b) display the snapshots of the transverse current and the electromagnetic field at the simulation time of 7.4 fs and 7.7 fs, respectively. $E_z + B_y$ represents the backward field while $E_z - B_y$ represents the forward field. One can see the transverse current in ES2 rises to peak in 0.3 fs due to the perturbation of AP1. If ES2 is at rest, such a transient current will lead to the emission of an electromagnetic pulse with a duration of about 300 attoseconds. However, the electron sheet is moving forward with a relativistic speed of $v_x$. The Doppler effect compresses the pulse in time domain by a factor of $4\gamma_x^2$, where $\gamma_x = 1/\sqrt{1-v_x^2/c^2}$ is the Lorentz factor. According to the simulation results, the $\gamma_x$ of ES2 is about 8. As a result, the duration of the generated half-cycle pulse is compressed to few-attosecond timescale in the lab frame. Figure 2c) illustrates the trajectories of representative electrons in ES2 from simulation time 7.2 fs to 8.1 fs. The emitted half-cycle pulse also can be viewed as the result of the synchrotron-like motion of the electrons.

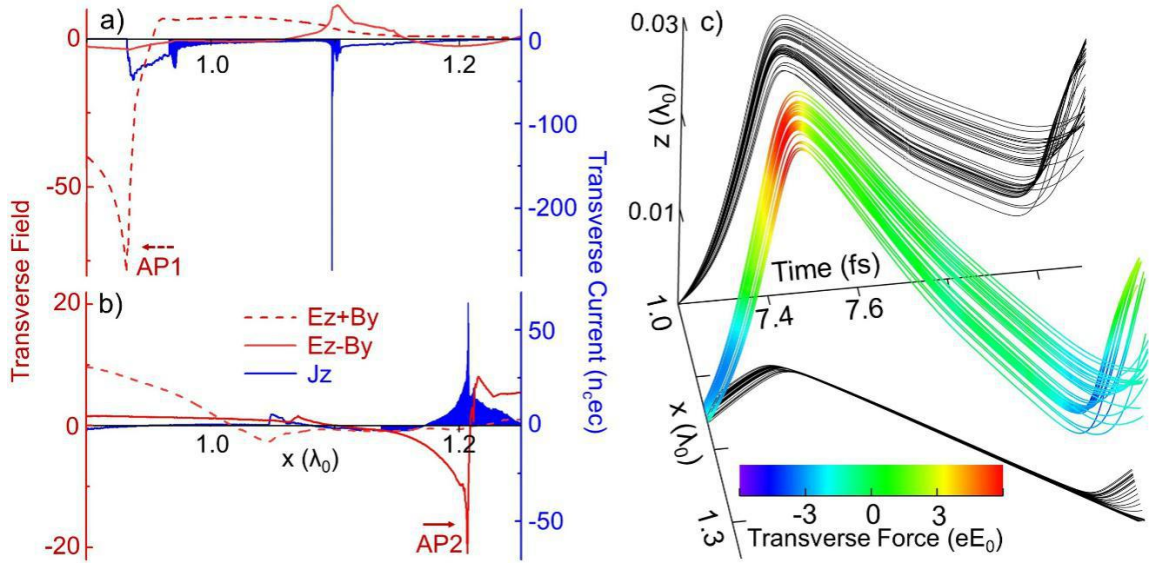

**Figure 2: The electron dynamics which determines the mission process of AP2.** a) Transverse current distribution (blue solid curve), and electromagnetic fields moving forward (red solid curve) as well as backward (red dashed curve) at the simulation time of 7.4 fs. b) The same as those shown in Fig. 2a) at the simulation time of 7.7 fs. The forward part of electrons is marked by the blue area. The transverse current is normalized by $j_0 = n_c ec$. The magnetic and electric field are normalized by $B_0 = m_e \omega/e$ and $E_0 = m_e \omega c/e$, respectively. c) The trajectories of representative electrons in ES2 from simulation time 7.2 fs to 8.1 fs. The colors of the lines represent the strength of transverse force normalized by $eE_0$. The black curves are the projections of the electron trajectories onto the x-t plane or z-t plane. The electrons' synchrotron motions starting from 7.4 fs leads to the emission of the half-cycle pulse in the forward direction.

**Robustness of the scheme**

The plasma density and the CEP of the driving pulse impose significant influences on the electron dynamics. We performed a series of simulations to study the robustness of our scheme by varying the plasma densities and the CEPs. As shown in fig. 3b) and 3c), for plasma densities of $35\, n_c < n_e <$

$43\, n_c$ and CEPs of $0.26 < \phi < 0.52$, the durations are all below 10 as. The achieved minimum pulse duration is 2.6 as, more than two orders of magnitude shorter than that of the driving pulse. At the meantime, the peak electric fields of the emitted half-cycle pulses keep at the same order of magnitude as the driving pulse.

To include the multi-dimensional effects, 2D PIC simulations were performed with parameters similar to that of 1D cases (see details in Methods). The electron dynamics and the emission process are pretty much the same. Figure 3d) displays one snapshot of the transverse field of the obtained attosecond pulse at the simulation time of 10 fs. As one can see, a half-cycle pulse with the duration of 8.3 as is generated. Considering that the numerical dispersion in 2D simulations would widen the pulse durations, maybe the actual durations could be even shorter in experiments. It is interesting to notice that the intensity of the obtained pulse in 2D simulations is higher than that in the 1D case due to the self-focusing effect[28].

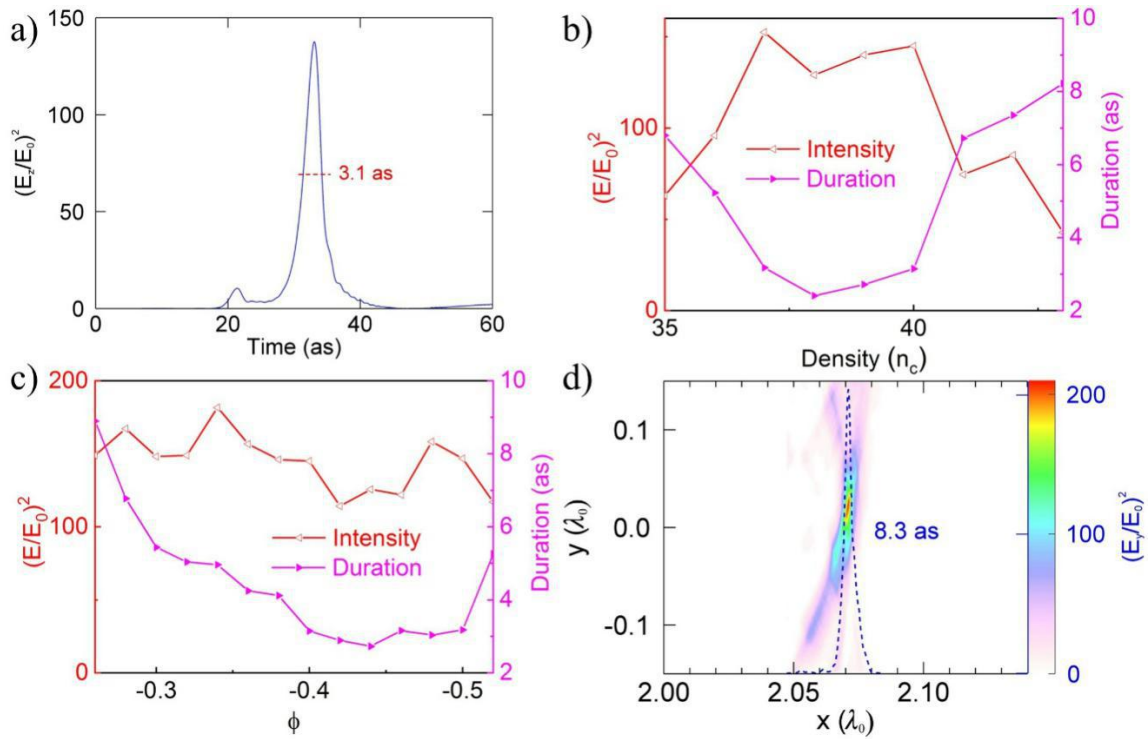

**Figure 3: 1D and 2D simulation results of the forward attosecond pulse.** a) The emitted half-cycle attosecond pulse at 1D simulation time of 9 fs. b) and c) The duration and normalized intensity $(E/E_0)^2$ of the attosecond pulse in dependence of the plasma densities and the CEPs φ of the one-cycle driving laser, respectively. d) One snapshot of the transverse field distribution of the obtained attosecond pulse at 2D simulation time of 10 fs.

**Realizing the scheme with a commercial 100-TW laser system**

Our scheme relies heavily on the pileup of the electrons in the plasma at first. The one-cycle pulse provides an ideal driving force for the electrons' pileup. As a matter of fact, our scheme also can be realized with multi-cycle pulses as long as they have a sharp rising edge to provide sufficient ponderomotive force. It has been demonstrated in theories and in experiments that a tens-of-femtosecond relativistic laser pulse can be shaped to a sharp-edged pulse by propagating through a NCD plasma[29]. Inspired by the related works, we perform 2D simulations by placing a NCD plasma slab in front of the overdense plasma to generate a half-cycle attosecond pulse. The driving laser is a 25-fs pulse with a normalized amplitude of a = 6 as displayed in Fig. 4a). Such a pulse can be produced with a commercial 100-TW laser system. The density and thickness of the NCD plasma were designed to achieve an optimal laser self-modification according to the plasma lens model[30]. As shown in Fig. 4b) and c), after a 25-$\lambda_0$-long NCD plasma, the front edge of the pulse is shaped highly similar to a one-cycle pulse. Then the cascaded generation of a half-cycle attosecond pulse happens in the overdense plasma. The intensity distribution of the emitted attosecond pulse after it propagating out of the plasma is displayed in Fig. 4d). The pulse duration is 46 as, which could be further reduced by utilizing a Petawatt femtosecond pulse and higher plasma density.

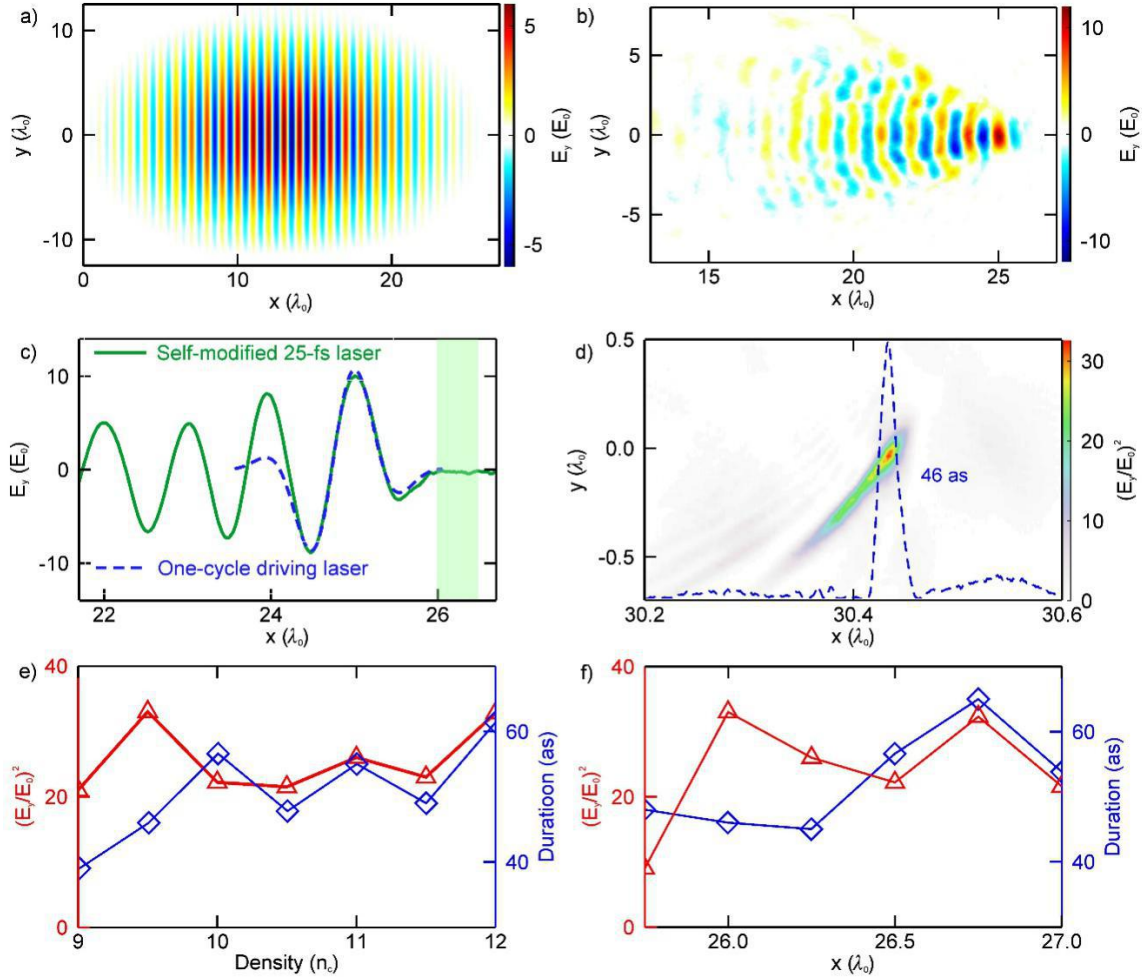

**Figure 4: Cascaded generation of an attosecond pulse in a double-layer target.** a) The initial multi-cycle driving laser with a duration of 25 fs and the normalized vector potential of a = 6. b) The shaped pulse with a sharp rising edge due to self-modulation and self-focusing effects after the laser propagating in the first-layer NCD plasma. c) The detailed profile of the one-cycle-like front edge at simulation time 42.5 $T_0$ (the same moment of Fig. 4b)). Here we display the mean value of electric field in the transverse range $-0.2\,\lambda_0 \leq y \leq 0.2\,\lambda_0$. d) The detailed intensity distribution of the emitted attosecond pulse propagating in the vacuum at simulation time 49.0 $T_0$. The blue dashed curve represents the distribution of intensity along the x-axis passing through the peak point of $E_y$. e) and f) The dependence of the duration (blue curve) and normalized intensity $(E_y/E_0)^2$ (red curve) of the attosecond pulse on the densities e) and locations f) of the second-layer target, respectively. The electric field is normalized by $E_0 = m_e \omega c/e$.

The robustness of the modified scheme is studied as well by varying the parameters of the targets. The influences of the second-layer plasma's densities are displayed in Fig 4e) indicating an allowable density fluctuation exceeding 10%. Figure 4f) illustrates the influences of the first-layer plasma's thicknesses on the attosecond pulse durations and amplitudes. Here the thicknesses of the first-layer NCD plasmas are determined by the locations of the second-layer target. The isolated half-cycle attosecond pulse can be generated in a wide range of thicknesses. By changing the thicknesses of the NCD plasma, the CEPs of the shaped pulses change accordingly (see details in Supplementary Information), which eventually determine the duration and amplitude of the generated attosecond pulse.

## Discussion

The key to getting a sub-10-attosecond half-cycle pulse in our scheme is to transversely perturb an ultrathin relativistic electron sheet with an attosecond pulse instead of a femtosecond laser pulse as in other schemes. The resulting pulse duration relies on the perturbation duration as well as the thickness and speed of the electron sheet. We developed a simple analytical model to include the above parameters to give an estimation of the pulse duration by taking into account the slowing-down of the electron sheet during emission.

We consider an ultrashort relativistic electron sheet with an initial speed of $v_0$ is perturbed by a counterpropagating electromagnetic field with constant amplitude of $E$ and $B$. For simplicity, assuming the electron sheet has a $\delta$-like density distribution and will not dilate. The duration of the emitted half-cycle pulse after $\Delta t$ is determined by the displacement of the electron sheet with respect to $c\Delta t$. If the electron sheet keeps its speed as a constant, the duration of the attosecond pulse will stretch linearly as $(c - v_0)\Delta t$. However, the existence of $\vec{v} \times \vec{B}$ force leads to the longitudinal deceleration of the electron sheet by $v(t) = v_0 - \frac{E^2 e^2 t^2}{m_0^2 \gamma_0^2 c}$. As a result, the duration of the emitted pulse $\tau \propto E^2 \Delta t^3$ (see details in Methods). In order to generate a sub-10-attosecond pulse, the perturbation time $\Delta t$ should be short enough, and the perturbation strength $E$ should not be too high. For example, if $\gamma_0 = 10$, $n_e = 1000\, n_c$, and $E = 2 \times 10^{13}$ V/m, the perturbation time should be less than 400 as. In our cascaded generation scheme, the electron sheet is perturbed by an attosecond pulse instead of a femtosecond pulse, which eventually leads to the generation of a sub-10-attosecond pulse. which is in agree with the simulation results in our cascaded generation scheme.

We use the parameters from simulation results as input to verify our model. The average $\gamma$-factor and maximum density $n$ of the electron sheet are 8 and $1050 n_c$, respectively. The electron distribution from simulation is applied instead of a $\delta$-like distribution (see details in Supplemental Material). The average field strength of driving pulse AP1 is estimated as $E = 8 \times 10^{12}$ V/m according to Figure 2a). Figure 5a) displays the evolution of transverse current density in the light-speed frame. One can see the deceleration of ES1 due to the driving pulse. Using the above parameters, the calculated duration and strength of field after a perturbation time $\Delta t$ is displayed in Fig. 5b), which are in agree with the PIC simulation results. It can be seen from Fig. 5b) that reducing the perturbation time will lead to a shorter attosecond pulse with the price of reduced intensity. It should be noted that there is a trade-off between the duration and intensity of the attosecond pulse by varying the perturbation time. We found the moderate overdense plasma is optimal in terms of generating a strong sub-10-attosecond pulse. For underdense plasmas, AP1 could not be generated due to the weak reflection of the driving laser pulse. For highly overdense plasma, the perturbation time will be too short to produce an attosecond pulse strong enough.

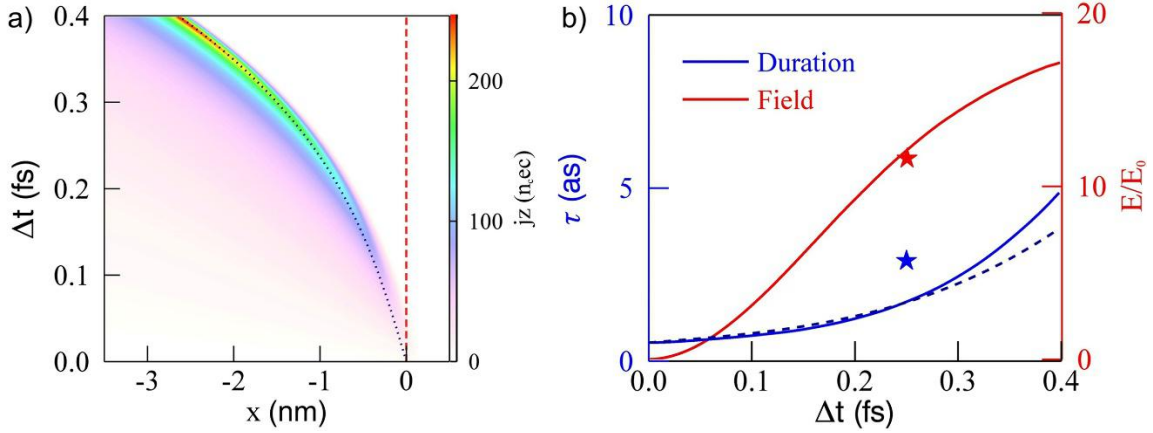

**Figure 5: The illustration of pulse stretching process.** a) The evolution of transverse current density $j_z$ at various perturbation time $\Delta t$ in the light speed frame. The red and blue curves present the locations of light and the electron sheet, respectively. The velocity of the electron sheet will decrease as $\Delta t$ increases. The emitted attosecond pulse will stretch as a result of the velocity variation between electron sheet and light. b) The dependence of the pulse duration $\tau$ and normalized field strength $E/E_0$ on the perturbation time $\Delta t$. The dashed line represents the case of a δ-like density distribution as Equation 3 in Method. The solid line shows the calculated duration (blue line) and strength of field (red line) respectively for the case of a realistic density distribution in simulations (see details in Supplementary Information). The stars represent the values generated directly from the simulations.

The experimental realization of the cascaded mechanism for an isolated half-cycle attosecond pulse requires the fabrication of targets and the characterization method of half-cycle attosecond pulses. Previous studies show carbon nanotube foams (CNF) can be utilized as the double-layer targets with the electron density ranges from 0.2 $n_c$ to 20 $n_c$. Such targets have been successfully applied for enhanced ion acceleration experiments[31]. Planar cryogenic hydrogen targets[32] or focused ion beam (FIB) processed plastic foil[33] can be used as the 40 $n_c$ targets. The techniques that directly measure the duration or waveform of the single-shot sub-10-attosecond pulses is highly challenging and may need a long time to be developed after the demonstration experiments. As the first step, a quick and reliable characterization is to measure the intensity scaling law of the spectra as compared with that of other schemes. We find the intensity scaling law of our scheme is $I(\omega) \propto \omega^{-8/3}$ (see details in Supplementary Information), different from the scaling of $\omega^{-4/3}$ or $\omega^{-6/5}$ from the coherent synchrotron emission (CSE) scheme[34,35]. The relativistically oscillating mirror (ROM) scheme can be excluded as it emits in the reflection direction[36]. The coherent wake emission (CWE) scheme has no contributions to the spectra above $7\omega_0$ due to its low cut-off energy as $\sqrt{n_e}\omega_0$[37]. Moreover, the attosecond pulse also can be confirmed by detecting the secondary effect resulted from the strong half-cycle electric field, for example, by setting a gas cell behind the overdense target and measuring the asymmetric electron spectra at different angles with respect to the laser axis[38].

In summary, we propose a novel scheme to generate intense isolated sub-10-attosecond half-cycle pulses in transmission direction based on a cascade process naturally happening in laser-plasma interaction. Systematic simulations reveal the robustness of this scheme. Presently, our scheme can be realized by shaping a 100-TW 25-fs laser pulse with plasma optics as the driving pulse. Such isolated pulses can be directly utilized as exceptional pump or probe pulses without the need for extra filters, offering new opportunities for high-field physics and ultrashort optics.

## Methods

### 1D Particle-in-cell simulations

The 1D and 2D simulations were carried out using the EPOCH code[39]. The incident one-cycle Gaussian laser is linearly polarized in z-direction with normalized vector potential $a = 30$. The temporal profile of the laser is $E_z = a \times exp(-(t-T_0)^2/\tau^2) \times sin(\omega t + \phi)$. Here $T_0$ and $\omega$ are the laser period and frequency with $\tau = 0.5 T_0$ while $\phi$ is the carrier envelope phase. Fully ionized sharp boundary plasma with a thickness of 0.5 λ and electron density of 40 $n_c$ is initially located between x = 1.0 λ and x = 1.5 λ, where λ = 800 nm is the laser wavelength and $n_c$ is the critical plasma density. The cell size is 40000 $cells/\lambda$ and each cell is filled with 100 macroparticles.

### 2D Particle-in-cell simulations

For the simulations of Fig 3d), the simulation box is 1λ × 8λ. The spatial resolution is λ/16000 × λ/200, and the number of particles per cell in the target is 16. The laser linearly polarizes in the simulation plane with a Gaussian transverse profile with a waist radius of $w_0 = 2\lambda$. The plasma is located at $0.1\lambda \leq x \leq 0.5\lambda$. All other parameters were the same as the 1D simulation above. Move window along x axis with the speed of $c$ is utilized from the simulation time of 2.2 $T_0$.

For the simulations of Fig. 4, the simulation box is 12.5λ × 25λ. The spatial resolution is λ/1000 × λ/40, and the number of particles per cell is 16. The laser linearly polarizes in the simulation plane with a Gaussian transverse profile with waist radius of $w_0 = 7.5$ λ. The first-layer NCD plasma is located at $1.0 \lambda \leq x \leq 26.0 \lambda$ with a density of 0.48 $n_c$. The second-layer overdense plasma is located at $26.0 \lambda \leq x \leq 26.5 \lambda$ with a density of 9.5 $n_c$. Move window is utilized which start move from the simulation time of 23.0 $T_0$ along x axis with light speed in vacuum.

### An analytical model predicting the duration of the half-cycle pulse

Considering an ultrashort relativistic electron sheet is transversely perturbed by a uniform electromagnetic field with strength $E$ starting from time $t = 0$. The forward radiation from the electron sheet is determined by the integral of the retarded transverse current as[34]

$$E(x,\Delta t) = \frac{1}{2\epsilon_0} \int_0^{\Delta t} j_\perp (x - c(\Delta t - t), t) dt \qquad (1)$$

Assuming the density profile of the electron sheet is unchanged during the emission process, the transverse current can be written as $j_\perp(x,t) = j(t)f[x - x_c(t)]$, where $f(x)$ is the local density function of the electron sheet, and $x_c(t)$ is the central position of the electron sheet. One can derive that $j_\perp = n_e e v_z = 2Ee^2 n_e t/m_0\gamma_0$, and $x_c(t) = v_0 t - 2E^2 e^2 t^3/3m_0^2\gamma_0^2 c$ (see details in Supplemental Material). Here $n_e$ is the plasma peak density, $\gamma_0$ is the initial relativistic factor of the electron sheet, and $v_0$ is the initial speed of the electron sheet. Finally, the field of the attosecond pulse generated from the corresponding transverse current after perturbation time $\Delta t$ can be expressed as

$$E(x,\Delta t) = \frac{1}{2\epsilon_0} \int_0^{\Delta t} \frac{2Ee^2 n_e}{m_0\gamma_0} t \, f(x - c\Delta t + (c - v_0)t + \frac{2E^2 e^2}{3m_0^2\gamma_0^2 c} t^3) dt \qquad (2)$$

For simplicity, assuming the electron density has a $\delta$-like distribution, then the duration of the attosecond pulse can be written as

$$\tau = (1 - \frac{\sqrt{2}}{2})(1 - v_0/c)\Delta t + (1 - \frac{\sqrt{2}}{4})\frac{2E^2 e^2}{3m_0^2 \gamma_0^2 c^2}\Delta t^3 \qquad (3)$$

## Acknowledgments

The PIC simulations were carried out in Shanghai Super Computation Center and High-performance Computing Platform of Peking University. This work is supported by National Basic Research Program of China (Grant No.2013CBA01502), National Natural Science Foundation of China (Grant Nos.11025523, J1103206, 11775010, 11535001, 61631001), NSFC innovation group project (11921006), The Project of Science and Technology on Plasma Physics Laboratory (No. 6142A04190111), Natural Science Foundation of Hunan Province (No. 2020JJ5031), and National Grand Instrument Project (2019YFF01014402).

## Conflict of interest

The authors declare no competing financial interests.

## Author contributions

Y.S. conceived the main idea and carried out all simulations, W.M. guided the illustration of the physics. W.M., X.Y., J.C. and G.M. conducted the work. Y.S. and R.H. developed the analytical model. Z.G clarifies some details of the physics. Y.S. and W.M. write the manuscript. All authors reviewed the manuscript.

# Supplementary Information of Cascaded Generation of a Sub-10-Attosecond Half-Cycle Pulse

Yinren Shou[1], Ronghao Hu[2], Zheng Gong[1], Jinqing Yu[3], Jia erh Chen[1], Gerard Mourou[4], Xueqing Yan[1,5], Wenjun Ma[1,*]

[1]State Key Laboratory of Nuclear Physics and Technology, and Key Laboratory of HEDP of the Ministry of Education, CAPT, Peking University, 100871 Beijing, China

[2]College of Physics, Sichuan University, 610065 Sichuan, China

[3]School of Physics and Electronics, Hunan University, 410082 Hunan, China

[4]DER-IZEST, Ecole Polytechnique, 91128 Palaiseau Cedex, France

[5]Collaborative Innovation Center of Extreme Optics, Shanxi University, 030006 Shanxi, China

[*]Corresponding authors. Email and telephone number: wenjun.ma@pku.edu.cn, +86-010-62760722 (W.J. M.)

The Supplementary Information contains three main parts. The first part presents the 1D PIC simulation results which are not included in the paper, electron dynamics derived for Method and the electron density function used in Fig. 5 in the paper. The second part illustrates the influence of the first-layer plasmas' thicknesses on the equivalent CEPs of the shaped driving laser in the double-layer targets scheme. The third part discusses some details for the experimental detections of the half-cycle attosecond pulse.

## 1D PIC Simulation Results Not Shown in the Paper

Figure s1 displays the shape of the one-cycle driving pulse utilized in the 1D PIC simulations. Here the word 'one-cycle' describes the driving laser's shape rather than its duration. Actually, the FWHM duration of the one-cycle driving pulse utilized in this work is 0.6 $T_0$.

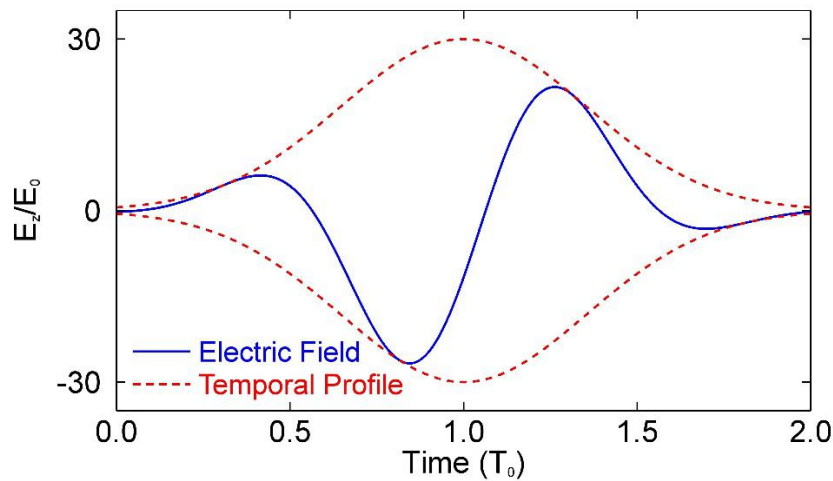

**Figure s1**: The shape of the one-cycle driving pulse utilized in the 1D PIC simulations.

Figure s2 illustrates the spatial-temporal intensity evolution of the forward field in the 1D PIC simulations. The duration of the generated forward half-cycle pulse (purple line) is compressed by more than two orders of magnitude as compared to that of the driving pulse (red line).

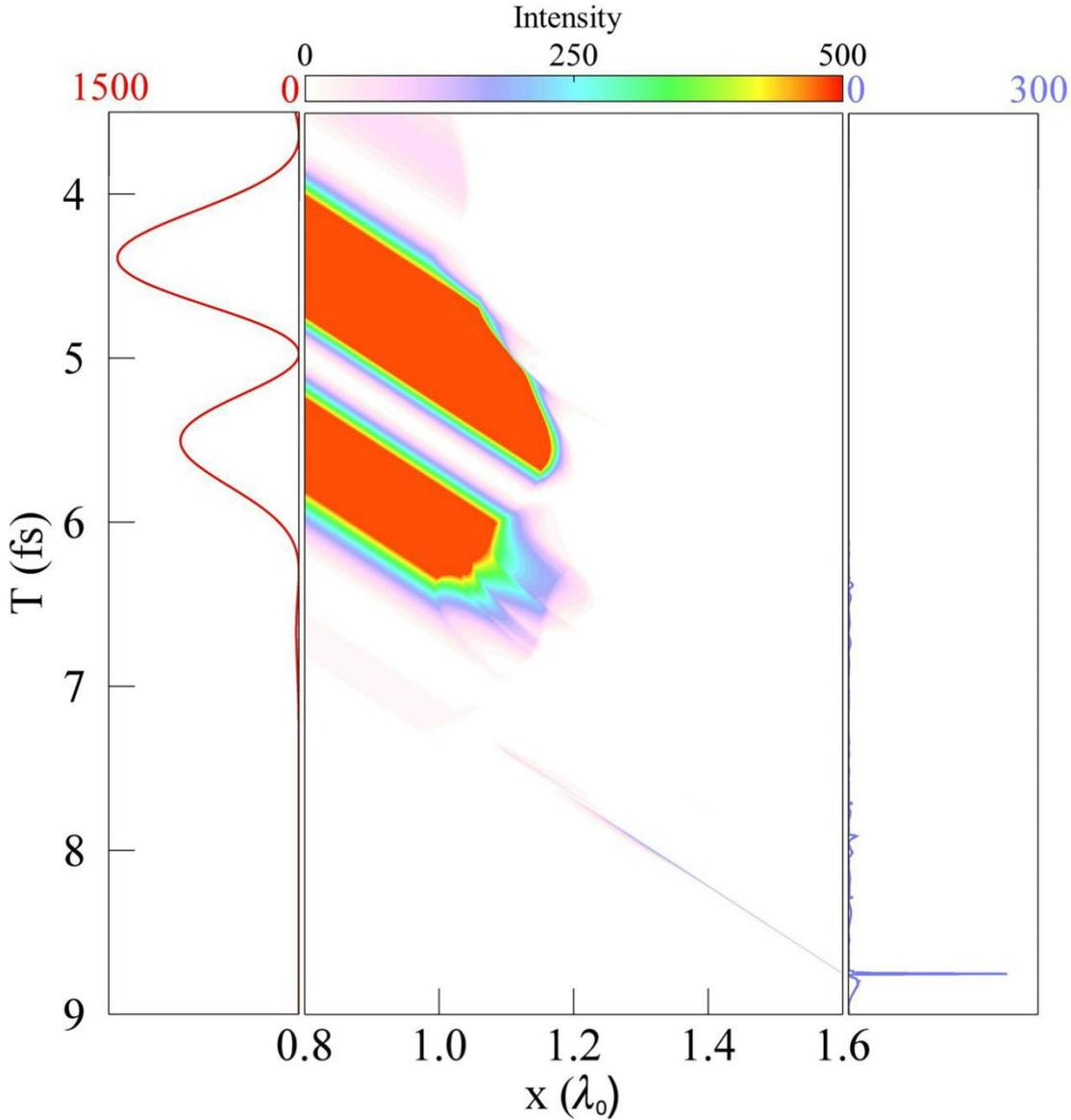

**Figure s2**: The spatial-temporal intensity distribution of forward field. Here the intensity is normalized as $(E_z/E_0 - B_y/B_0)^2/2$.

The cascaded scheme can be achieved even if the boundary of the target is not perfectly sharp. We introduce preplasmas with density profiles of $n_e = n_0 exp(-x/L)$ at the boundary of the plasma in simulations. Figure s3 depicts the dependence of duration $\tau$ and normalized intensity $(E/E_0)^2$ of AP2 on the parameter $L$. One can see half-cycle attosecond pulses with the durations less than 10 as can be obtained with $L \leq 0.05 \lambda_0$, which can be achieved in experiments with high-contrast laser pulses.

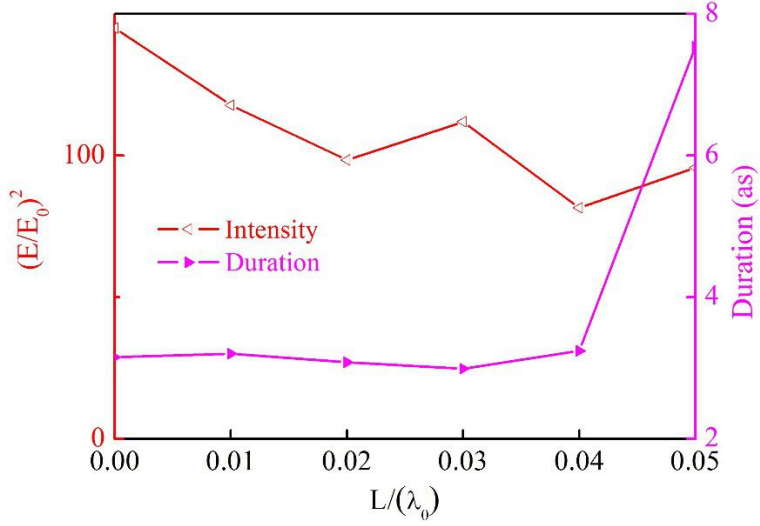

**Figure s3**: The dependence of the duration τ and normalized intensity $(E/E_0)^2$ of the half-cycle attosecond pulse on the scale length L of the preplasma.

## Electron Dynamics

When an electron sheet with relativistic factor $\gamma \gg 1$ is perturbed by a constant electromagnetic field with amplitude of $E$ and $B$, assuming $v_z(0) = 0$, the transverse dynamics can be expressed as

$$m_0 \frac{d(\gamma v_z)}{dt} = F_z = -(E + Bv_x)e \approx -2Ee \qquad (1)$$

In the first order approximation, $v_z$ can be written as

$$v_z \approx \frac{-2Ee}{m_0 \gamma_0} t \qquad (2)$$

The derivative of $\gamma$ can be expressed as

$$\frac{d\gamma}{dt} = \frac{-Eev_z}{m_0 c^2} = \frac{2E^2 e^2}{\gamma_0 m_0^2 c^2} t dt \qquad (3)$$

Then the longitudinal dynamics also can be expressed as

$$m_0 \frac{d(\gamma v_x)}{dt} = F_x = \frac{-2BEe^2}{m_0 \gamma_0} t \qquad (4)$$

As a result, the longitudinal position of the sheet is

$$x = v_0 t - \frac{E^2 e^2}{m_0^2 \gamma_0^2 c} \frac{t^3}{3} \qquad (5)$$

## Electron Distribution

The realistic density distribution in simulations is displayed in Figure s3. Here we use the density distribution of ES2 at simulation time 7.6 fs considering the fact that the sub-10 attosecond pulse is emitted during simulation time 7.4 fs to 7.7 fs. The density profile can be approximately expressed as

$$n_a(x) = \begin{cases} c + \frac{3}{2a^3 \ln(x_{max}/x_{min})} \times [(x-a)\{x+(x+a)[\frac{1}{2}\ln(\frac{x_{min}}{a-x})]\} \\ \quad + x_{min}(\frac{1}{2}x_{min} + 2x)], & x \in [-a - x_{min}, a - x_{min}] \\ \\ c + \frac{3}{2a^3 \ln(x_{max}/x_{min})} \times [-2ax - (x^2 - a^2)\ln(\frac{a-x}{-a-x})], & x < -a - x_{min} \\ \\ c, & x > a - x_{min} \end{cases} \quad (6)$$

The parameters for Fig. s3 can be obtained from the simulations as $x_{min} = 2 \times 10^{-4}$, $x_{max} = 0.2$, $a = 1 \times 10^{-4}$, $c = 60$.

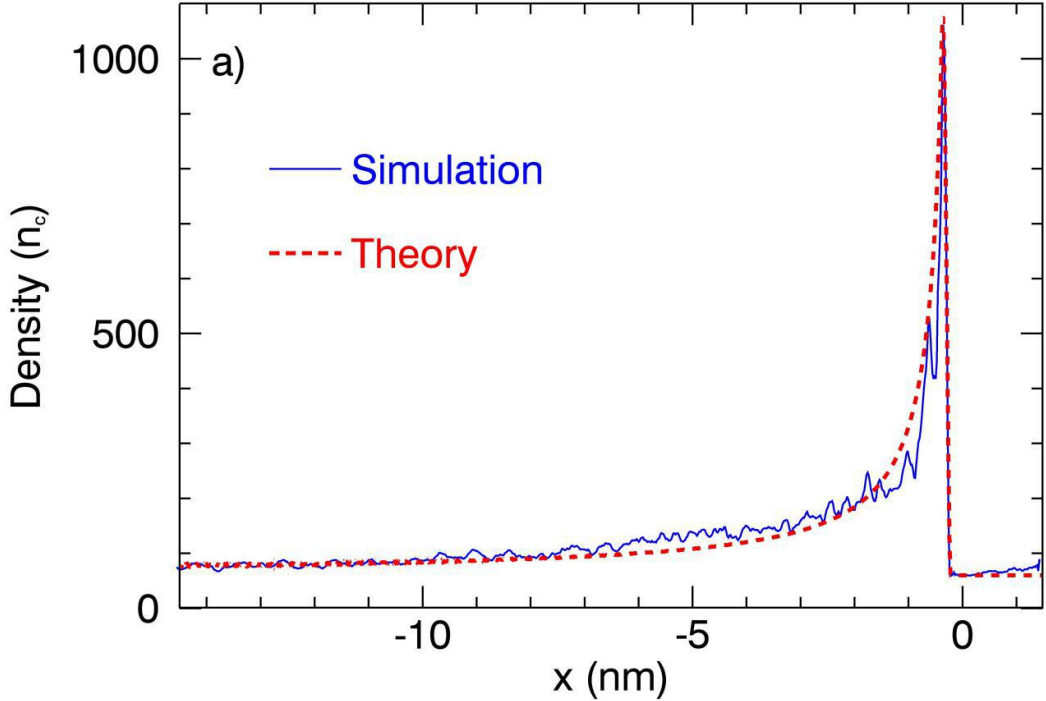

**Figure s3**: The electron density distribution of ES2 at simulation time 7.6 fs. The blue line represents the simulative result while the red dashed line is the electron density profile calculated according to Eq. 6.

## CEPs of the Equivalent One-Cycle Pulse for the Shaped Laser

The change of second-target locations corresponds to a modification in CEPs of the equivalent one-cycle pulse as displayed in Fig. s5. The detailed parameters of the one-cycle pulses are list in the legend. Here the thicknesses of the first-layer NCD plasmas are determined by the locations of the second-layer target.

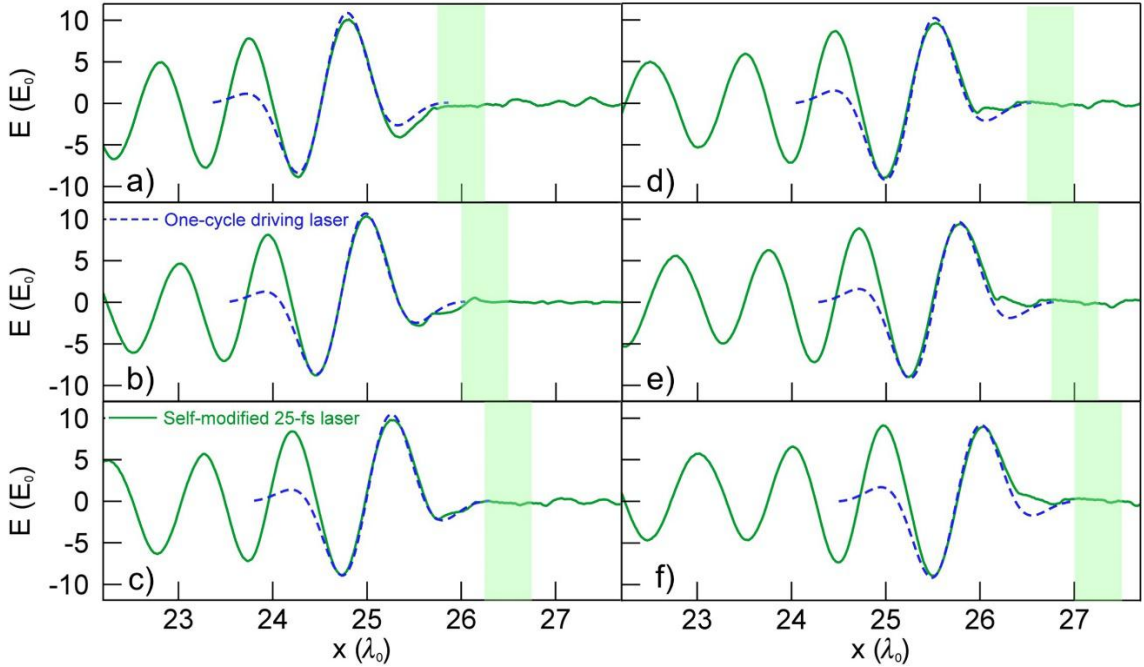

**Figure s5**: Carrier envelope phases (CEPs) of the equivalent one-cycle pulse corresponding to varied second-target locations. The normalized vector potentials of the equivalent one-cycle pulses for a) - f) are 12.0, 12.0, 12.0, 12.0, 11.6, 11.3, while the CEPs are -0.5, -0.4, -0.3, -0.2, -0.1, 0.0, respectively. The yellow block indicates the location of the second-layer target.

## Experimental Detections

Figure s6 displays the experimental detections of the generated attosecond pulse utilizing a 25-fs driving laser. The decay scaling of intensity on the harmonic order are $n^{-8/3}$, similar to the relativistically oscillating mirror (ROM) scheme, and varying from $n^{-4/3}$ or $n^{-6/5}$ of coherent synchrotron emission (CSE) scheme. Considering that ROM is in the reflection directions while we detect the XUV spectra in the transmission direction, the $n^{-8/3}$ scaling law can distinguish the attosecond pulse emitted from the cascaded mechanism or CSE scheme. Another detection method is to compare the electron spectra measured in 30° and -30° for the double-layer and the triple-layer targets. Here a triple-layer target consists a double-layer target and an additional 3 μm thick, 0.1 $n_c$ plasma appended behind the second-layer target. The half-cycle structure of the attosecond pulse can be verified through the asymmetric angular distribution of accelerated electrons as shown in Fig s6b).

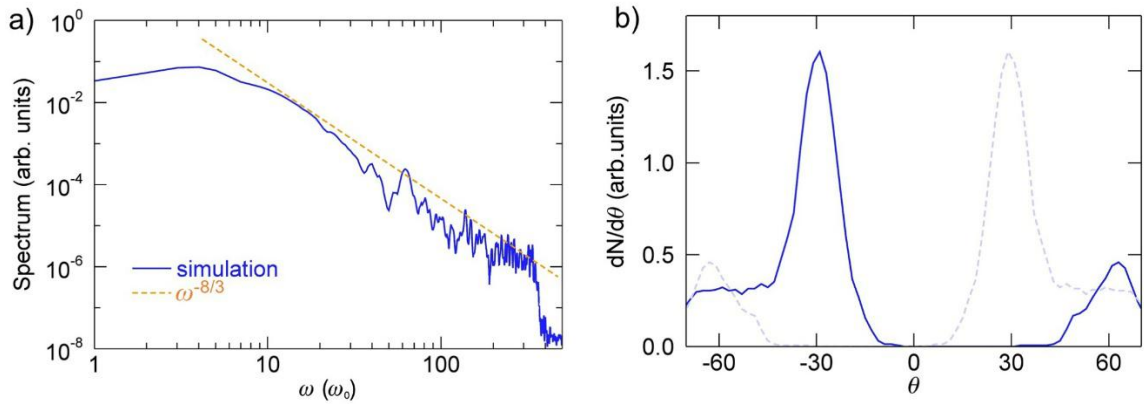

**Figure s6**: Experimental detections of the generated half-cycle attosecond pulse. A flat-field spectrometer can be utilized to obtain XUV spectra at the laser propagation direction. The decay scaling of intensity on the harmonic order are $n^{-8/3}$ as shown in a). b) the asymmetric angular distribution of accelerated electrons from the third-layer target. Here we count the high-energetic electrons with energies greater than 1 MeV.